\documentclass[manuscript]{aastex}
\usepackage{graphicx}
\usepackage{txfonts}
\usepackage{natbib}

\shorttitle{Limb-darkened Radiation-Driven Stellar Winds}
\shortauthors{Cur\'e et al.}

\begin{document}
\title{Limb-Darkened Radiation-Driven Winds from Massive Stars}
\author{M. Cur\'e} 
\affil{Departamento de F\'{\i}sica y Astronom\'{\i}a, Facultad de Ciencias, 
Universidad de Valpara\'{\i}so\\
Av. Gran Breta\~na 1111, Casilla 5030, Valpara\'{\i}so, Chile}
\email{michel.cure@uv.cl}
\and
\author{L. Cidale\altaffilmark{1}}
 \affil{Departamento de Espectroscop\'{\i}a, Facultad de Ciencias 
 Astron\'omicas y Geof\'{\i}sicas, 
 Universidad Nacional de La Plata (UNLP), 
and\\ Instituto de Astrof\'{\i}sica La Plata, CCT La Plata, CONICET-UNLP\\
Paseo del Bosque S/N, 1900 La Plata, Argentina}
\email{lydia@fcaglp.unlp.edu.ar}
\and
\author{D. F. Rial\altaffilmark{1}}
\affil{Departamento de Matem\'{a}ticas, Facultad de Ciencias 
Exactas y Naturales, \\
Universidad de Buenos Aires, Argentina.
\and
IMAS, CONICET}
\email{drial@dm.uba.ar}

\altaffiltext{1}{Member of the Carrera del Investigador Cient\'{\i}fico, 
CONICET, Argentina}

 \begin{abstract}
We calculated the influence of the limb-darkened finite disk correction factor 
in the theory of radiation-driven winds from massive stars.
We solved the 1-D m-CAK hydrodynamical equation of rotating radiation-driven 
winds for all three known solutions, i.e., fast, $\Omega$-slow and $\delta$-slow.
We found that for the fast solution, the mass loss rate is increased
by a factor 
$\sim 10\%$, while the terminal velocity is reduced about $10\%$, when compared 
with the solution using a finite disk correction factor from a
uniformly bright star.
For the other two slow solutions the changes are almost negligible. 
Although, we found that the limb darkening has no effects on the
wind momentum luminosity relationship, it would affect  
the calculation of synthetic line profiles and the derivation of 
accurate wind parameters.
\end{abstract}

\keywords{hydrodynamics --- methods: analytical--- stars: early-type --- stars: mass-loss
--- stars: rotation --- stars: winds, outflows} 
%
%
\section{Introduction}

The CAK theory \citep{cas75} describes the mass loss due to radiation force in massive stars. 
This theory is based on a simple parameterization of the line force ($\alpha$ and $k$) 
which represents the contribution of the spectral lines to the radiative acceleration by a power 
law distribution function. \citet{abb82} improved this theory calculating the line force considering 
the contribution of the strengths of the hundreds of thousands of lines. He also included a 
third parameter ($\delta$) that takes into account the change in ionization throughout the 
wind. Despite this immense effort to give a more realistic representation of the line force, 
evident discrepancies still remained. Further improvements to this theory done by \citet{fri86} 
and \citet{pau86} (hereafter m-CAK model) relaxed the point star 
approximation with the introduction of the finite disk correction
factor, assuming a uniform bright spherical source of
radiation. From then on, this model has succeeded in describing
both, wind terminal velocities ($v_{\infty}$) 
and mass-loss rates ($\dot{M}$) from very massive stars. 
As a result of the radiation force, the properties of the stellar winds must somehow reflect 
the luminosities of the stars. This relationship can be obtained from the line driven wind theory 
\citep{kud95,kud99} and, nowadays, it is known as the Wind Momentum--Luminosity 
Relationship (WM--L).
It predicts a strong dependence of wind momentum rate on the stellar luminosity with  
$\alpha$ \citep{pul96}.

The  m-CAK hydrodynamical solution (hereafter the fast solution) is characterized by an exponential 
growth at the base of the wind that matches very quickly a $\beta$-law profile when the velocity 
reaches some few kilometers per second, with a $\beta$ index in the range
0.8 to 1.0.
  
However, in the last decade, \citet{cur04} and \citet{cur11} found two new physical 
solutions from the 1-D non-linear m-CAK hydrodynamics equation that describe the wind 
velocity profile and mass loss rates from rapidly rotating stars (the $\Omega$-slow solution) 
and from slowly rotating  A- and late B-type supergiants (the $\delta$-slow solution). 
The $\Omega$-slow solution only exists when the star's rotational speed is larger than $\sim$ 
3/4 of the breakup speed. This $\Omega$-slow solution posses a
larger mass loss rate (the higher the rotational speed, the higher the mass loss rate) 
and reaches a terminal velocity which is about 1/3 of the fast solution's terminal speed. 
On the other hand, the $\delta$-slow solution is found  when the line-force parameter 
$\delta$ is slightly larger than $\sim$ 0.25. High values of $\delta$ are expected in hydrogen 
rich environments; for a pure hydrogen gas \citet{pul00} demonstrated that  $\delta$ is $1/3$.
This last solution, where the Abbott $\delta$ factor represents changes in the 
ionization of the wind with distance, reaches a slow terminal velocity, similar to the 
$\Omega$-slow solution, but with a much lower mass loss rate. 

In the m-CAK model, the calculation of the radiation force  is often carried
out assuming a uniform bright finite-sized spherical star. 
The rapid rotation, however, changes the shape of the 
star to an oblate configuration  \citep{co95,pel00} and induces
gravity darkening \citep{vz24} as function of (co)-latitude. In both cases, 
a rotating and non-rotating star, the decrease of the temperature
outwards the photosphere produces a limb darkening effect which also
modifies the finite disk correction factor. The theoretical formalism
for computing the self-consistent radiation
force for non-spherical rotating stars, including the effects of stellar
oblateness, limb darkening and gravity darkening,  was  developed by
\citet{co95}. However, to disentangle the effects of each one of these
competing processes upon the wind structure, these authors present a
semi-quantitative analysis and estimated that the
limb darkening effect could increase the mass loss rate ($\dot{M}$)
in an amount of  $\sim 11\%$ to $\sim 13\%$ over the uniformly bright models.
However, that larger mass loss would imply a reduction in the wind terminal speed.  
\citet{oud04} carried out (for the fast solution) a perturbation
analysis of the effects of the gas pressure on the mass loss rate and
wind terminal velocity in terms of the ratio of sound speed to escape speed 
($a/v_{\mathrm{esc}}$). They showed that for finite-disk-corrected 
spherical wind, typical increases in mass-loss rate are 10\%--20\%, with 
comparable relative decreases in the wind terminal speed. 

Then, considering that the radiative flux does not change significantly when
the limb darkening is taken into account, an enhancement of $\sim 10\%$  
in the mass loss rate might lead not only to a lower  terminal speed ($v_{\infty}$) 
but also to a change in the theoretical WM--L.  An accurate determination of 
the WM-L relationship for A and B supergiants (Asgs and Bsgs) is important 
because it would allow the use of these stars as extragalactic distance 
indicators \citep{bre04}. 

In this work, we present an analytical expression for the
limb darkening finite disk correction factor and solve the 1-D
hydrodynamical equation for all three known solutions for radiation
driven winds; i.e., fast, $\Omega$-slow and $\delta$-slow solutions.
These results are compared with the wind solutions
computed with the finite disk correction factor assuming  
a uniform bright star, finding that the effects of the limb darkening 
are only important for fast solution.

In  \S 2 we briefly describe the 1-D momentum equation of the wind,
in \S 3 we present an analytical expression for the 
limb-darkened finite disk correction factor and in \S 4 we solve
numerically the  hydrodynamics equations for model parameters
corresponding to the fast, $\Omega$-slow,  $\delta$-slow  
and $\Omega\delta$-slow solutions.
Finally, in \S 5, we discuss the results, conclusions
and future work.

\section{The m-CAK hydrodynamic model \label{HYD}} 
The m-CAK model for radiation driven winds considers one dimensional component 
isothermal fluid in a stationary regime with spherical symmetry. 
Neglecting the effects of viscosity, heat conduction  and magnetic fields 
\citep{cas75}, the equations of mass conservation and radial momentum read:  
\begin{equation} 
4\pi\, r^{2}\rho\, v=\dot{M},  \label{2.0} 
\end{equation} 

\noindent{and}
  
\begin{equation} 
v\,\frac{dv}{dr}=-\frac{1}{\rho }\frac{{dp}}{dr}-\frac{GM\,(1-\Gamma )}{r^{2}}+ 
\frac{v_{\phi }^{2}(r)}{r}+g^{line}\,(\rho,dv/dr,n_{E}).  \label{2.1}
\end{equation} 

\noindent Here $v$ is the fluid velocity and $dv/dr$ its gradient. All other variables have their
standard meaning (see \citet{cur04} for a detailed derivation and definitions of variables, 
constants and functions). We adopted the standard
parametrization for the line force  term, given by \citet{abb82}, \citet{fri86}, \citet{pau86}:
\begin{equation} 
g^{line}=\frac{C}{r^{2}}\;f_{\mathrm{\,D}}\,(r,v,dv/dr)\;\left( r^{2}\,v\,\frac{dv}{dr} 
\right) ^{\alpha }\;\left( \frac{n_{E}}{W(r)}\right) ^{\delta }  \label{2.2},
\end{equation} 
where the coefficient $C$ depends on $\dot{M}$, $W(r)$ is the dilution factor
and $f_{\mathrm{\,D}}$ is the finite disk correction factor. \\

Introducing the following change of variables  
$u =-R_{\ast }/r$, $\,w =v/a$ \,and $\,w'=dw/du$, where $a$ is the isothermal 
sound speed and $a_{\mathrm{rot}}=v_{\mathrm{rot}}/a$, where $v_{\mathrm{rot}}$ 
is the equatorial rotation speed at the stellar surface, the momentum equation becomes: 
\begin{equation}
F(u,w,w^\prime) \equiv \left( 1-\frac{1}{w^{2}} \right)
w\,\frac{dw}{du}+A+\frac{2}{u}+a_{\mathrm{rot}}^{2}\,u-C^{\prime } 
\;f_{\mathrm{\,D}}\;g(u)\,(w)^{-\delta }\left( w\,\frac{dw}{du}\right) ^{\alpha 
}\ = 0. 
\label{2.5} 
\end{equation}

The standard method for solving this non-linear differential equation
(\ref{2.5}) together with the constant $C^{\prime }(\dot{M})$ (eigenvalue of this
problem) is imposing that the solution passes through a singular (or critical)
point.\\
Critical points are defined at the roots of the singularity condition, namely:
\begin{equation} 
\frac{\partial}{\partial w^{\prime}}\,F(u,w,w^{\prime} 
)=0.  \label{2.6} 
\end{equation} 
At this specific point and in order to find a physical wind solution, a regularity condition 
must be also imposed, i.e.,
\begin{equation} 
\frac{d }{du}\,F(u,w,w^{\prime}) =\frac{\partial F}{\partial u}+\frac{\partial 
F}{\partial w}\,w^{\prime}=0.  \label{2.7} 
\end{equation} 

In order to solve this equation we need to know the behaviour of the finite disk 
correction factor $f_{\mathrm{\,D}}$. To disentangle limb darkening from rotational effects 
(gravity-darkening and oblateness) we will analize them independently. A  discussion on the effects 
of the oblate finite disk correction factor on the velocity profile and mass loss rate was 
presented by \citet{ara11}. Therefore, in this work we mainly discuss the importance of the 
limb darkening on radiation driven winds.
 
\section{Limb-darkened finite disk correction factor}

\citet{co95} derived an integral expression for the limb-darkened finite disk 
correction factor ($f_{\mathrm{\,LD}}$), based in a simple linear gray atmosphere, namely:
\begin{equation} 
f_{\mathrm{\,LD}}\,(r,v,dv/dr)=\frac{r^2}{R_{\ast}^2 \,(1+\sigma)^\alpha} \int_{\mu_{\ast}}^{1} 
(1+\sigma \mu^{\prime 2})^\alpha 
\times \left(1 + \frac{3}{2} \sqrt{\,\frac{\mu^{\prime 2} -\mu_{\ast}^2}{1-\mu_{\ast}^2}} \right) 
\,\mu^\prime d\mu^\prime,
\label{2.8} 
\end{equation}
where $\sigma \equiv (d\ln{v}/d\ln{r})-1$\, and \,$\mu_{\ast}=\sqrt{1-R_{\ast}^2/r^2}$.\\
The integration of eq. \ref{2.8} gives the following analytical expression :
\begin{eqnarray}
 f_{\mathrm{\,LD}}\,(r,v,dv/dr)&=& \frac{1}{2\, \sigma\, (\alpha +1)} \times
\left(\frac{\sigma
   +1}{-\frac{\sigma
   }{r^2}+\sigma
   +1}\right)^{\alpha } \times   \nonumber \\ 
& & \left[\, \sigma \,
(\alpha +1)
   \; _2\- F_1\left(\frac{3}{2
   },-\alpha
   ,\frac{5}{2},-\frac{\sigma
   }{r^2\, (\sigma +1)-\sigma
   }\right) \right. \nonumber \\
   & &\left.+r^2\,(\sigma +1)
   \left(\left(\frac{\sigma
   +1}{-\frac{\sigma
   }{r^2}    +\sigma
   +1}\right)^{\alpha
   }-1\right)+\sigma  \right]
   \label{2.9} 
 \end{eqnarray}
where $_2 F_1$ is the Gauss Hypergeometric function.

\begin{figure}[ht]
\epsscale{1.}
\plottwo{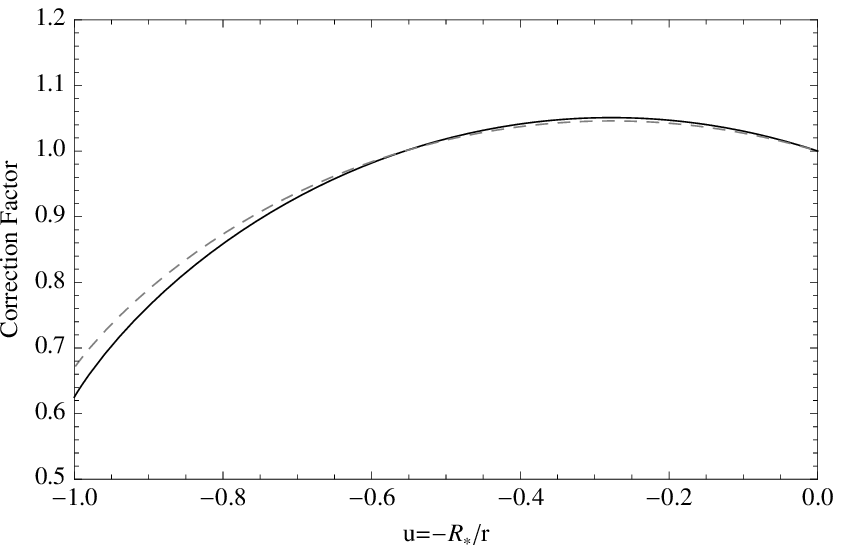}{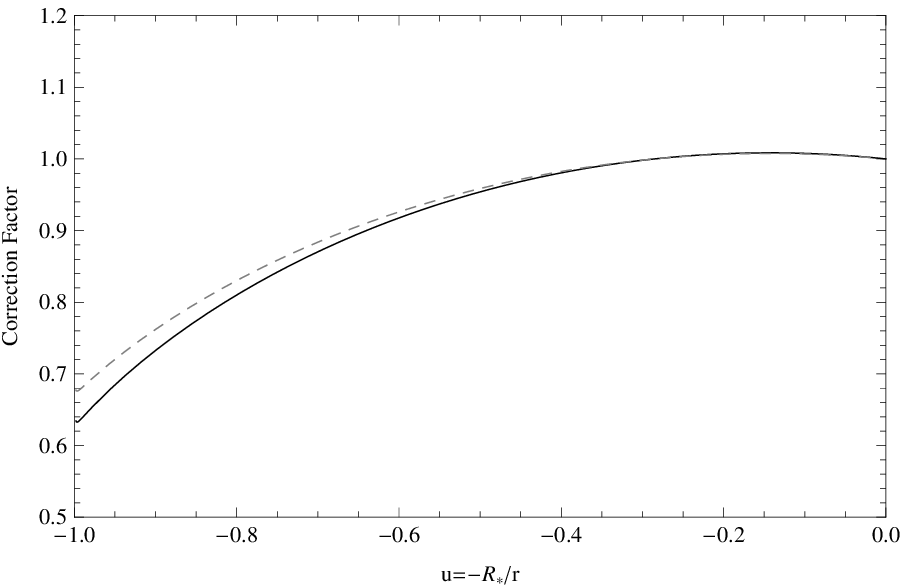}
\caption{
{\small{Uniformly bright $f_D$ (continuous line) and limb-darkened $f_{LD}$ (gray-dashed line) finite 
disk correction factors. Calculations where performed using a $\beta$-law velocity field with 
$\beta=0.8$ (left panel) and $\beta=2.5$ (right panel). In both cases the value for $\alpha$ was 0.6}
} \label{fig1}}
\end{figure}

Figure \ref{fig1} compares the run of both uniformly bright and 
limb-darkened finite disk correction factors as function $u$, 
using two different $\beta$-law velocity profiles ($\beta$ = 0.8 and 2.5) 
and a typical value of the $\alpha$ line-force parameter equals 0.6. This figure 
clearly shows that at the base of the wind the factor $f_{\mathrm{\,LD}}$ 
(shown in gray-dashed line) is about $\sim 10\%$ larger than the one obtained 
for a uniform bright stellar disk, $f_{\mathrm{\,D}}$  (continuous line), 
increasing, the value of the mass loss rate. Instead, at larger distances 
from the stellar surface, both correction factors have the same behaviour 
as a function of $u$.

Mathematically, $f_{\mathrm{\,LD}}$ can be considered as a small perturbation of 
$f_{\mathrm{\,D}}$, as it is shown in Figure \ref{fig1}, even for different $\beta$-law index.
Thus, based on the standard theory of dynamical system \citep[see, e.g.,][]{PDM82}, 
we expect no large differences when considering velocity profiles from the equation of motion (eq. \ref{2.5}), 
for the cases where uniformly bright or limb–-darkened finite disk correction factors are used. 
This is a consequence of the theorem of the continuous dependence of the solutions of the ordinary 
differential equations (ODE) on their parameters \citep{HS74}.

Therefore, the scope of this paper is limited to the study of the numerical 1-D stationary solutions 
of eq. \ref{2.5}, leaving a theoretical topological analysis (and also a time--dependent one)
for a future work.

\section{Results \label{results} }
We are now in conditions to solve the non-linear differential equation (eq. \ref{2.5}) 
considering the factor $f_{\mathrm{\,LD}}$ given by eq. \ref{2.9}. 
The calculation of all partial derivatives of $f_{\mathrm{\,LD}}\,(u,w,w')$ are 
given in Appendix \ref{A}.  These derivatives are needed in order to evaluate 
the singularity and regularity conditions (Eqs. \ref{2.6} and \ref{2.7}, respectively).
Depending on the selected parameter-space, each type of solution explains the wind of 
a different kind of massive object, i.e., fast solution describes the wind of hot stars, 
the $\Omega$-slow solution explains the wind of rapid rotators such as Be stars, and the 
$\delta$-slow solution characterizes the wind of A-type supergiants.  
In the following subsections we will adopt a prototype star for each one of these three 
known physical solutions, in order to analyse the effects of the limb darkening on 
the m--CAK hydrodynamical model.

\subsection{Fast Solution}
\begin{figure}[ht]
\epsscale{1.}
\plottwo{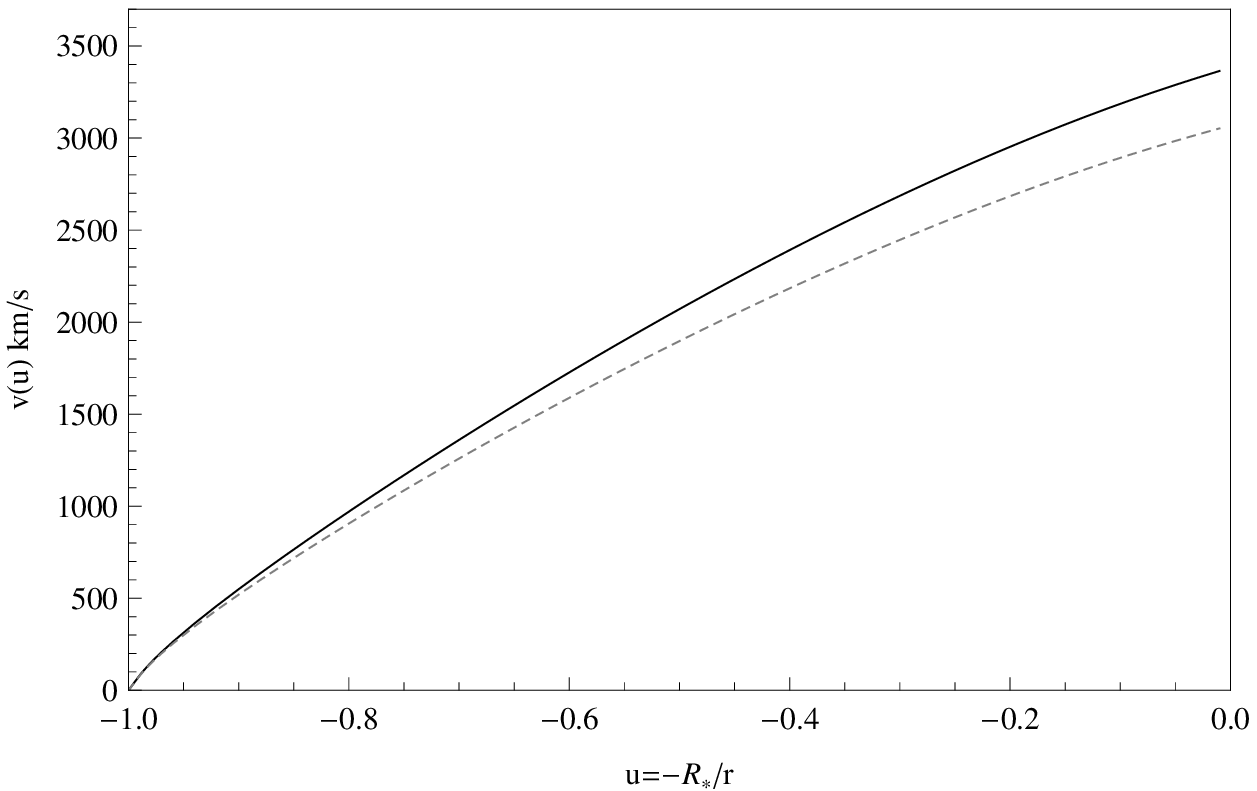}{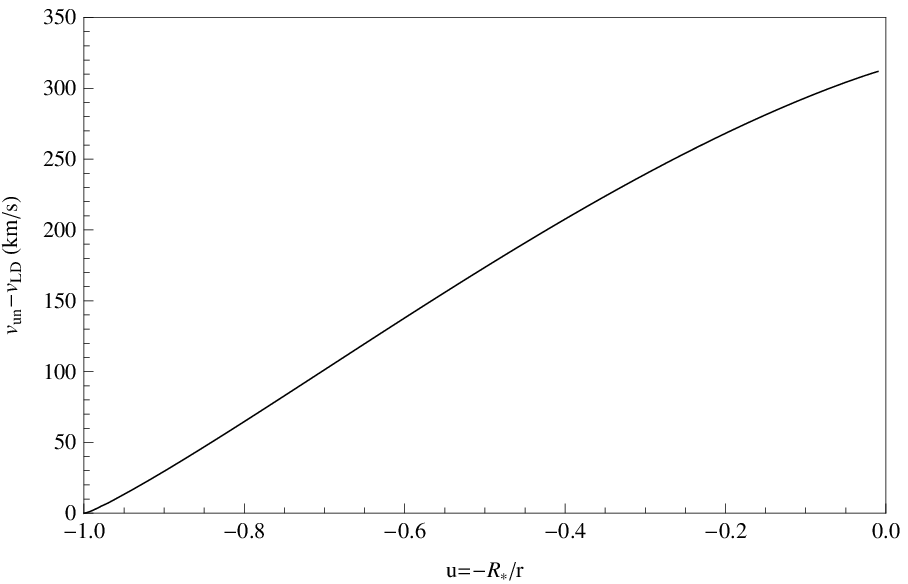}
\caption{Left panel: velocity profile as function of the inverse radial coordinate $u$. 
The standard m-CAK model is shown in continuous line
and the solution with the limb-darkened finite disk correction factor is in gray dashed line. 
The effect of the $f_{\mathrm{\,LD}}$ in the velocity profile 
is very significant reducing the terminal velocity approximately in $10\%$ with respect to 
the standard m-CAK model. Right panel: Velocity difference
between the standard  solutions with a uniformly bright, $f_{\mathrm{\,D}}$, 
and the limb-darkened, $f_{\mathrm{\,LD}}$, correction factors.
{\small{}
} \label{fig2}}
\end{figure}

For the standard fast solution we selected, as in \citet{cur04}, a typical O5\,V star 
with the following stellar and line-force parameters: $T_{\mathrm{eff}}$ = 45\,000 K, 
$\log\,g$ = 4.0, $R/R_{\sun}$ = 12, $v_{\mathrm{rot}}$ = 0, $k$ = 0.124, $\alpha$ = 0.64 
and $\delta$ = 0.07 \citep{lam99}. 
The numerical code we used to solve the momentum equation is described in 
\citet{cur04}. Figure \ref{fig2} (left panel) shows the velocity profile for the standard 
case, where a uniform bright star disk (continuous line) and the limb-darkened
one (gray-dashed line) are used.  Figure \ref{fig2} (right panel) displays the difference 
in the velocity, $\Delta v = v_{\mathrm{un}} - v_{\mathrm{LD}}$ (where $v_{\mathrm{un}}$ 
is the velocity profile when the uniform finite disk correction factor is taken 
into account, while $v_{\mathrm{LD}}$ is the wind solution obtained using $f_{\mathrm{\,LD}}$). 
The $v_{\mathrm{\,LD}}(u)$ profile is always smaller than the $v_{\mathrm{un}}(u)$ one, with 
a monotonically increasing  difference. The effect of the limb-darkened finite disk correction 
factor changes the behaviour of the velocity field  in the most external layers, it 
reaches a smaller terminal velocity by about $10\%$ of the $v_{\infty}$ value of the 
standard m-CAK case. 
There is no significant change of the velocity field at the base of the wind. Therefore, 
the location of the singular point is almost the same in both cases. Our calculations 
confirmed the predictions of \citet{co95,oud04}  based on the behaviour of the $f_{\mathrm{\,LD}}$ 
at the base of the wind, i.e., that the mass loss rate is increased a factor of about 
$\sim 10\%$. Concerning to the WM-L relationship, the value of 
$D_{\mathrm{mom}}=(\dot{M}\, v_{\infty} \sqrt{R_{\ast}/R_{\sun}})$ shows almost 
no change due to a compensation of the increase in the mass loss rate  and the decrease of the terminal 
velocity, as shown in Table \ref{tab1}. Although the value of
$D_{\mathrm{mom}}$ seems to remain unaltered, we would expect minor
differences in the synthetic spectra when they are computed with the two different 
velocity profiles. 

\begin{table*}[t!]
\begin{center}
  \caption{{\small{Wind parameters for the fast solution with uniformly bright ($f_{\mathrm{\,D}}$) and 
  limb-darkened ($f_{\mathrm{\,LD}}$) finite disk correction factors}}}
  \label{tab1}
 {\scriptsize
  \begin{tabular}{lcc}
\hline
\hline
  &$f_{\mathrm{\,D}}$&$f_{\mathrm{\,LD}}$\\
\hline
$\dot{M}$ ($10^{-6}$ $M_{\sun}\,yr^{-1}$)					& 2.206				& 2.449				\\
$v_{\infty}$ ($km\, s^{-1}$)									& 3384				& 3071  				\\
$r_{\rm{singular}}$  $(R_{\ast})$ 							& 1.027				& 1.031 				\\
EigenValue ($C^{\prime}$)									& 40.89				& 38.53				\\
$\log\,D_{\mathrm{mom}}$ (cgs) 								& 29.21				& 29.22				\\
\hline
  \end{tabular}
  }
\end{center}
\end{table*}

These results show that the correction to the line radiation force
due to the limb darkening effect leads to lower mass loss rates and
higher  wind terminal velocities, both in approximately 10\%, when
compare with the contribution of a uniformly bright star disk radiation source.

\subsection{$\Omega$-Slow Solution}
\label{o-slow}
The $\Omega$-slow  solution is present when the star is rotating at
velocities near the breakup rotational speed. Therefore, 
to study the effects of the limb darkening in the radiation force we
select, the case of a typical B1\,V star 
with high rotational speed ($\Omega$ = $v_{\mathrm{rot}}/v_{\mathrm{breakup}}$
= 0.9)  and the following stellar parameters:
$T_{\mathrm{eff}}$ = 25\,000 K, $\log\,g$ = 4.03, $R/R_{\sun}$ = 5.3.
The corresponding line force parameters: $k$ = 0.3, $\alpha$ = 0.5 and
$\delta$ = 0.07 were taken from \citet{abb82}. 

\begin{figure}[ht]
\epsscale{1.}
\plottwo{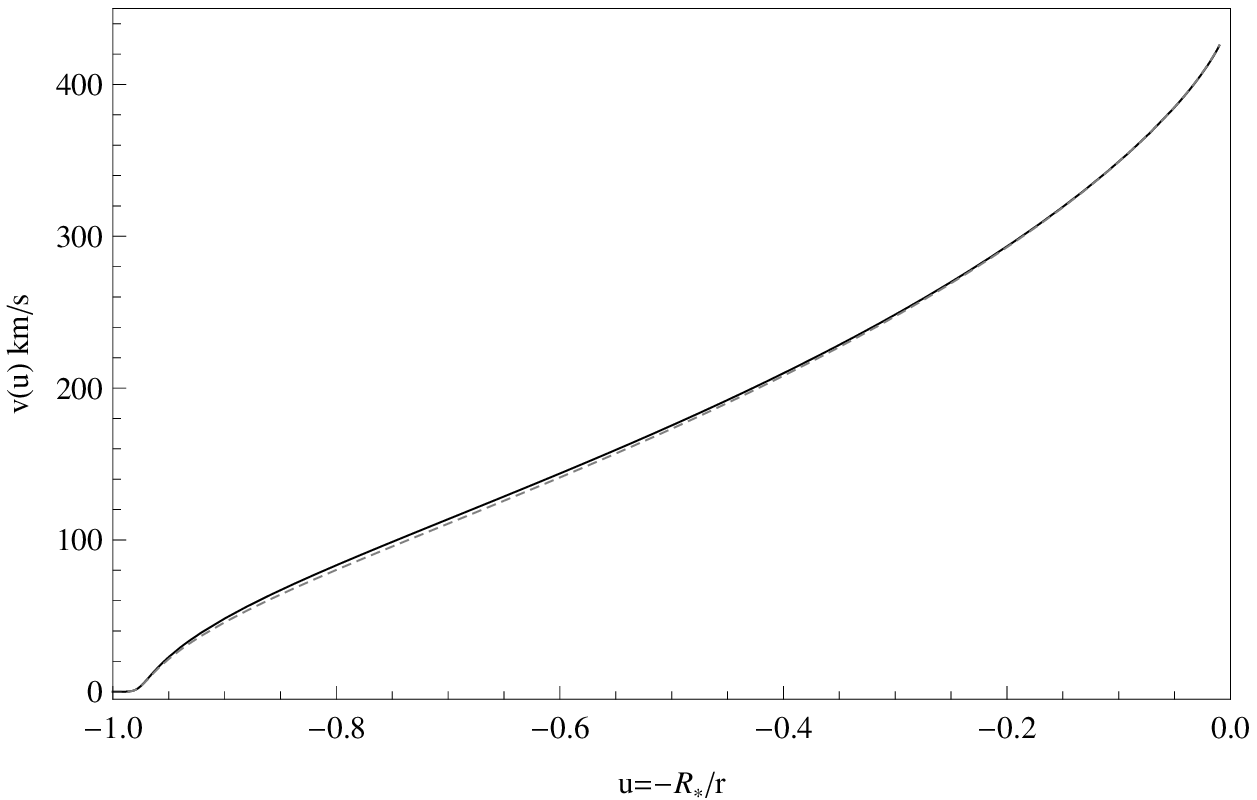}{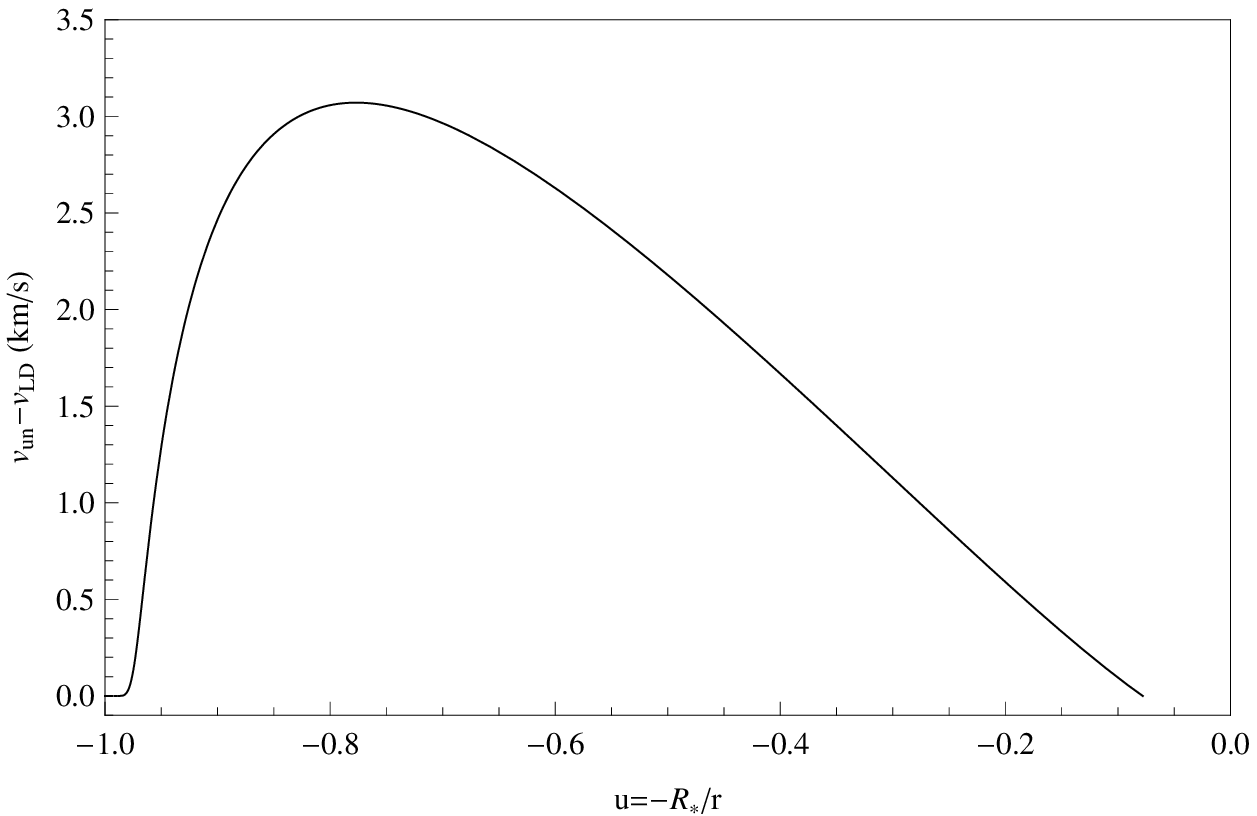}
\caption{Left panel: Same as Figure \ref{fig1},  $v(u)$ versus $u$. In this case the 
effect of the $f_{\mathrm{\,LD}}$ in the velocity profile is minimal. Right panel: Velocity difference.
{\small{}
} \label{fig4}}
\end{figure}

\begin{table*}[ht!]
\begin{center}
  \caption{{\small{Wind parameters for the $\Omega$-slow solution with uniformly bright ($f_{\mathrm{\,D}}$)
  and limb-darkened ($f_{\mathrm{\,LD}}$) finite disk correction factors}}}
  \label{tab3}
 {\scriptsize
  \begin{tabular}{lcc}
\hline
\hline
  &$f_{\mathrm{\,D}}$&$f_{\mathrm{\,LD}}$\\
\hline
$\dot{M}$ ($10^{-6}$ $M_{\sun}\,yr^{-1}$)	& 4.22 $10^{-3}$	& 4.22 $10^{-3}$	\\
$v_{\infty}$ ($km\,s^{-1}$)					& 446.8		& 446.5				\\
$r_{\rm{singular}}$  $(R_{\ast})$ 			& 26.14		& 26.14 				\\
EigenValue ($C^{\prime}$)					& 78.31		& 78.27				\\
$\log D_{\mathrm{mom}}$ (cgs) 				& 25.44	    	& 25.44				\\
\hline
  \end{tabular}
  }
\end{center}
\end{table*}
The resulting velocity profiles with uniform and limb-darkened
correction factors and the corresponding differences in the velocities 
are shown in Figure \ref{fig4}. These plots show
clearly that the effect of $f_{\mathrm{\,LD}}$ in the velocity
profile is minima. The influence of the $f_{\mathrm{\,LD}}$ on the mass loss rate 
and other wind quantities are shown in Table \ref{tab3}, together with the comparison 
of the velocity profile using the uniform correction factor. All the changes in these 
quantities are minimal or even negligible. 
There is an important dominance of the centrifugal force term in the momentum equation (\ref{2.5}).

\subsection{$\delta$-Slow Solution}
For the calculation of the $f_{\mathrm{\,LD}}$ correction factor in  the
parameter-space of the 
$\delta$-slow solution, we select an  A-type supergiant
star with the following fundamental parameters: 
$T_{\mathrm{eff}}$ = 10\,000 K, $\log\,g$ = 2.0, $R/R_{\sun}$ = 60, $v_{\mathrm{rot}}$ = 0, 
and line-force parameters: $k$ = 0.37,
$\alpha$ = 0.49 and $\delta$ =0.3 \citep[model W03 from][]{cur11}. 

\begin{figure}[ht]
\epsscale{1.}
\plottwo{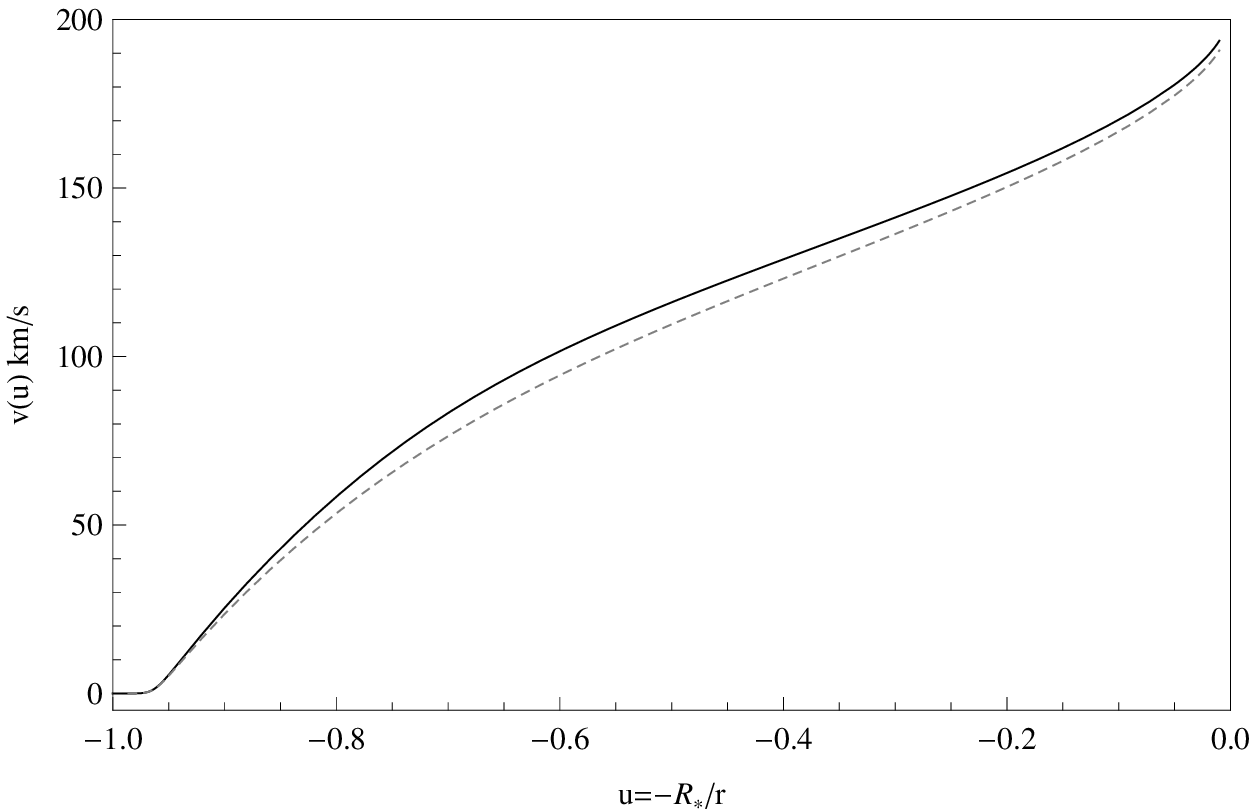}{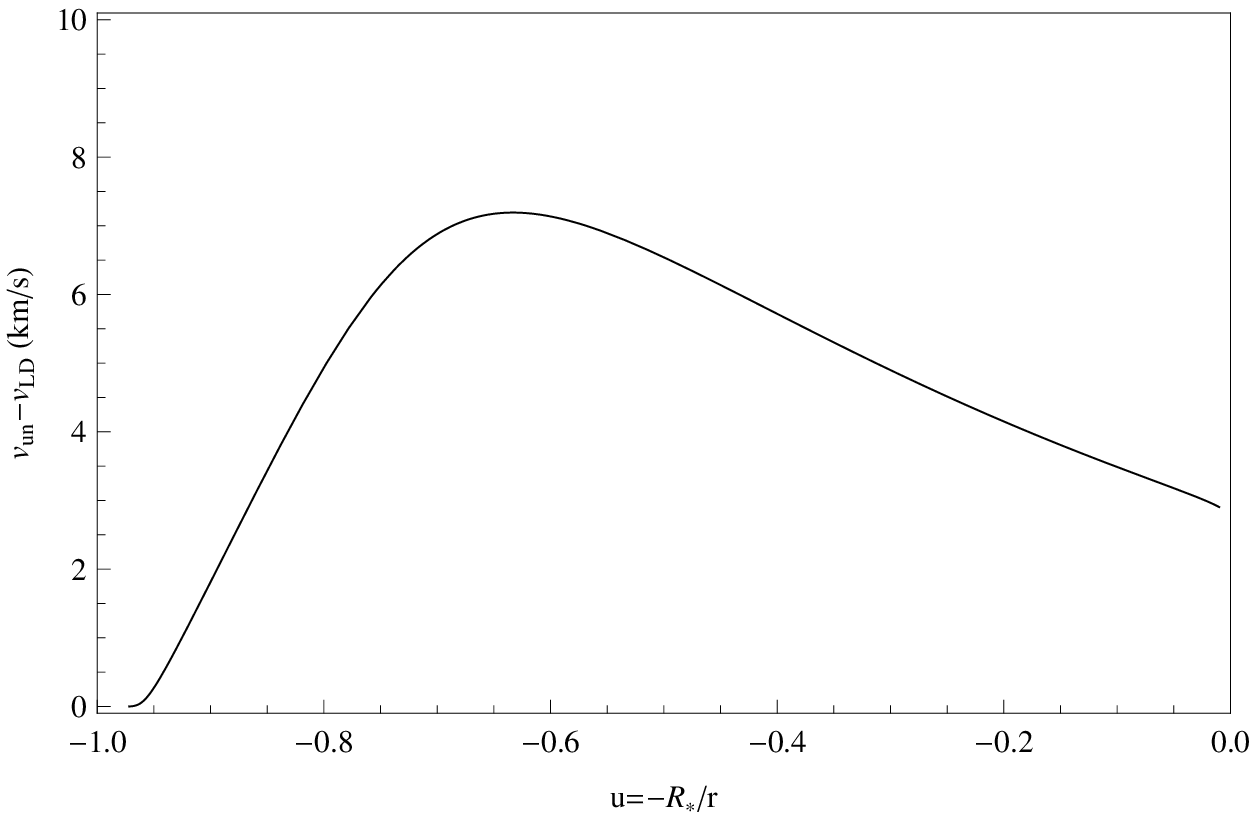}
\caption{Left panel: Same as Figure \ref{fig1}, $v(u)$ versus $u$. In this case the effect of 
the $f_{\mathrm{\,LD}}$ in the velocity profile is minimal. Right panel: Velocity difference.
{\small{}
} \label{fig3}}
\end{figure}

Similar to the $\Omega$-slow wind solution, the effect of the
limb darkening is negligible in both, the velocity profile and mass
loss rate (see Figure \ref{fig3} and Table \ref{tab2}).

Concerning the influence of the limb darkening on the WM-L relationship, 
there is no substantial effect.

\begin{table*}[ht!]
\begin{center}
  \caption{{\small{Wind parameters for the $\delta$-slow solution with uniformly bright ($f_{\mathrm{\,D}}$)  
  and limb-darkened ($f_{\mathrm{\,LD}}$) finite disk correction factors}}}
  \label{tab2}
 {\scriptsize
  \begin{tabular}{lcc}
\hline
\hline
  &$f_{\mathrm{\,D}}$&$f_{\mathrm{\,LD}}$\\
\hline
$\dot{M}$ ($10^{-6}$ $M_{\sun}\, yr^{-1}$)					& 7.22 $10^{-4}$	 	& 7.36 $10^{-4}$		\\
$v_{\infty}$ ($km\,s^{-1}$)									& 203				& 200				\\
$r_{\rm{singular}}$  $(R_{\ast})$ 								& 11.06				& 11.06 				\\
EigenValue ($C^{\prime}$)									& 63.78				& 63.54				\\
$\log\, D_{\mathrm{mom}}$ (cgs)			& 24.85				& 24.86				\\
\hline
  \end{tabular}
  }
\end{center}
\end{table*}

\subsection{$\Omega \delta$-Slow Solution}
\begin{figure}[ht]
\epsscale{1.}
\plottwo{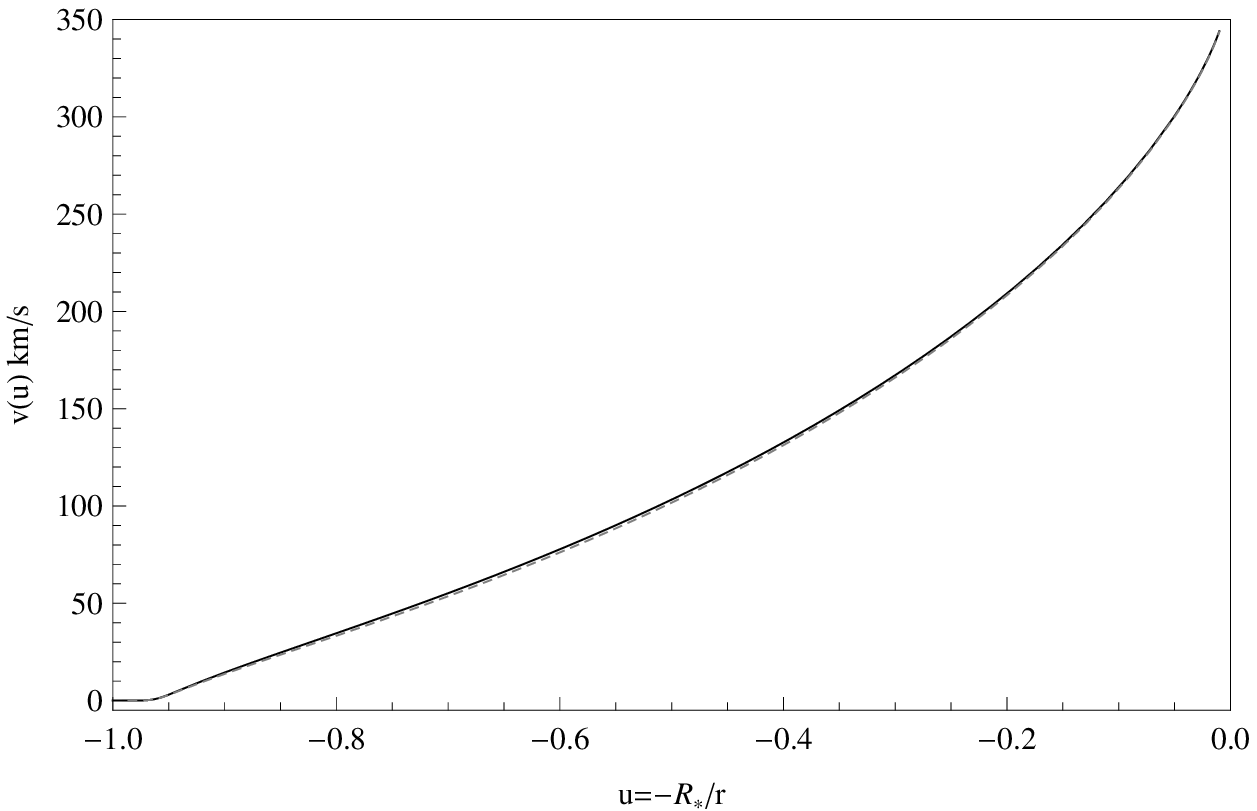}{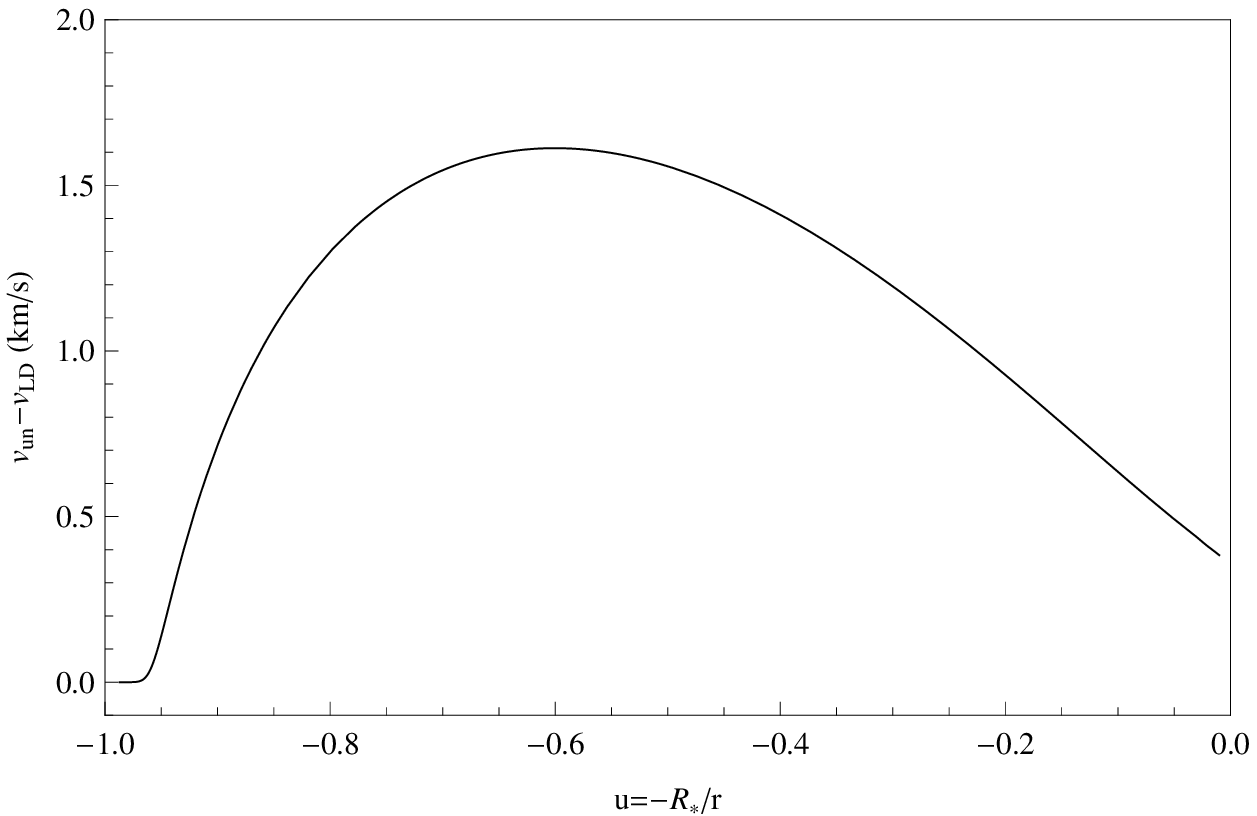}
\caption{Left panel: Same as Figure \ref{fig1},  $v(u)$ versus $u$. In this case the 
effect of the $f_{\mathrm{\,LD}}$ in the velocity profile 
is minimal. Right panel: Velocity difference.
{\small{}
} \label{fig5}}
\end{figure}

Here we investigate the particular case when $\Omega$  and $\delta$
take higher values. We selected the same test star as in \S \ref{o-slow} 
but with a different value of the $\delta$ parameter 
($\delta$ = 0.25).  The computed hydrodynamic solutions for uniformly bright  
and limb-darkened correction factors are almost the same, as it shown 
in Figure \ref{fig5} and Table \ref{tab4}.

\begin{table*}[ht!]
\begin{center}
  \caption{{\small{Wind parameters for the $\Omega \delta$-slow solution with uniformly 
  bright ($f_{\mathrm{\,D}}$) and  limb-darkened ($f_{\mathrm{\,LD}}$) finite disk correction factors}}}
  \label{tab4}
 {\scriptsize
  \begin{tabular}{lcc}
\hline
\hline
  &$f_{\mathrm{\,D}}$&$f_{\mathrm{\,LD}}$\\
\hline
$\dot{M}$ ($10^{-6}$ $M_{\sun}\,yr^{-1}$)			& 8.63 $10^{-4}$	    & 8.64 $10^{-4}$		\\
$v_{\infty}$ ($km\, s^{-1}$)									& 367.8				& 367.5			  			\\
$r_{\rm{singular}}$  $(R_{\ast})$ 							& 38.17				& 38.17 						\\
EigenValue ($C^{\prime}$)									& 113.9				& 113.9						\\
$\log\, D_{\mathrm{mom}}$ (cgs)						& 24.66				& 24.66						\\
\hline
  \end{tabular}
  }
\end{center}
\end{table*}

When we compare the velocity profiles between the $\Omega$-slow solution 
computed with $\delta$ = 0.07 (see figure \ref{fig3} left panel)
and $\delta$ = 0.25 (see figure \ref{fig5} left panel), we find that both profiles have the 
same behaviour as a function of $r$. Therefore, the centrifugal term due to the high dominates 
over the $\delta$-factor in $g^{line}$.
Nevertheless, the influence of the $\delta$-factor is not negligible, it reduces the mass loss rate
in $\sim 80\%$ and the terminal velocity in $\sim 20\%$.

\section{Discussion and Conclusions}
\begin{figure}[ht]
\epsscale{1.}
\plottwo{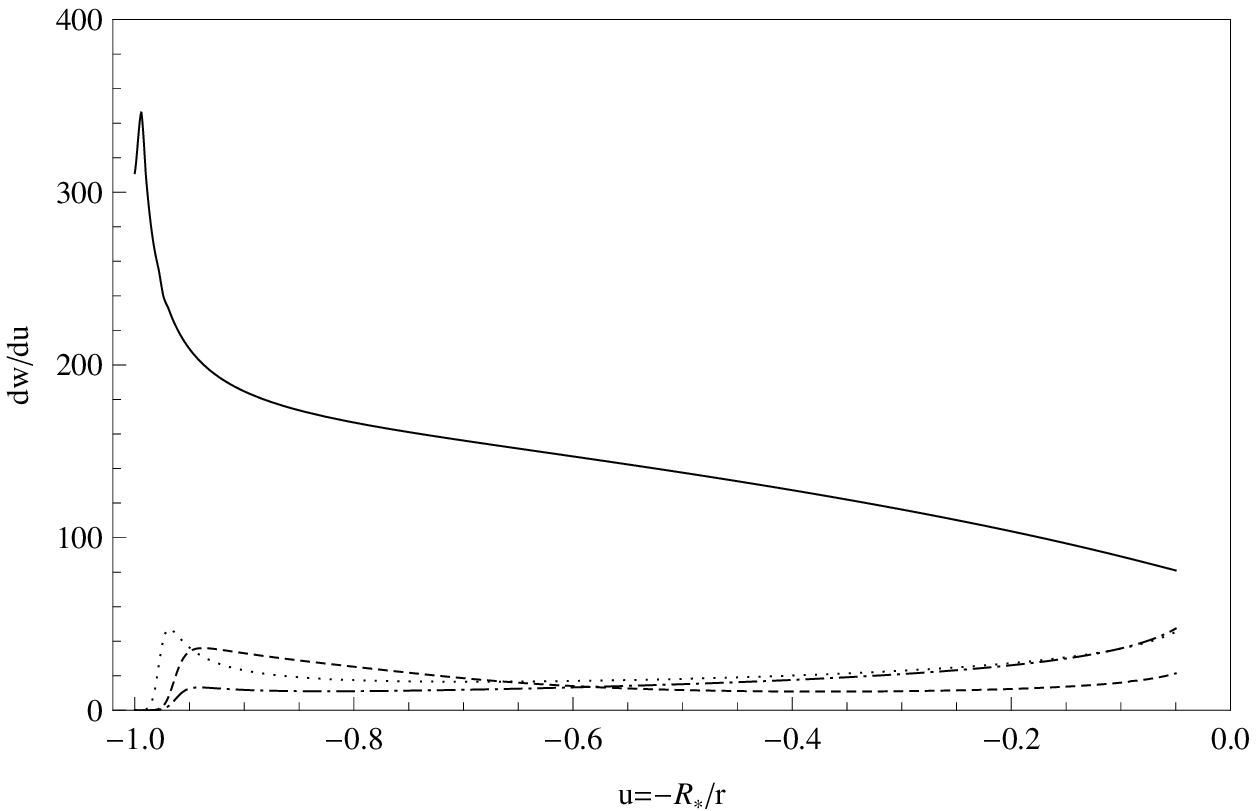}{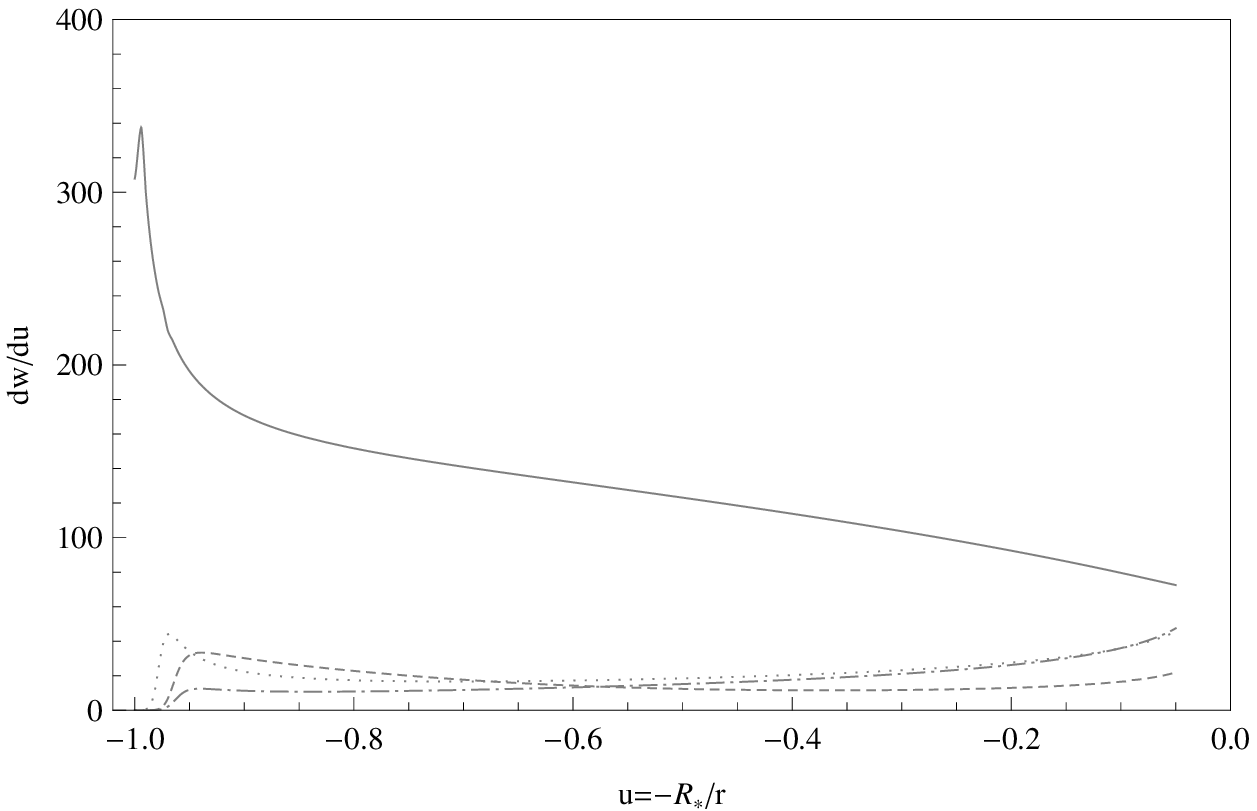}
\caption{Left panel: Normalized velocity gradient $dw/du$ versus $u$ from the solutions 
of the equation of motion using the uniform finite disk $f_D$. In continuous-line  
is plotted the gradient of the fast solution; dashed--line correspond to the $\delta$-slow solution;
dotted--line to the $\Omega$-slow solution and dashed--dotted--line to the $\Omega \delta$-slow solution.
Right panel: id, for the cases where the $f_{LD}$ us used in the equation of motion (eq. \ref{2.5})
{\small{}
} \label{fig6}}
\end{figure}

In this work we improved the description of the radiation
force taking into account the correction factor due to  a
limb-darkened disk. In particular, we derived an analytical formula
to compute this contribution. Then, we solved the 1-D non-linear momentum 
equation for radiation driven winds and analized the influence of 
$f_{\mathrm{\,LD}}$ for all the three known solutions, namely: the fast, the
$\Omega$-slow and the $\delta$-slow solutions, as well as, the case
  of a high $\Omega$ and $\delta$ parameter, the $\Omega$ $\delta$-slow solution.

We selected the appropriate  stellar parameters of massive
stars that are representative of each possible hydrodynamical
solution and evaluated the  velocity profile as function of the radial coordinate.

We found a significant impact of $f_{\mathrm{\,LD}}$ in the radiation driven-wind 
of massive stars that are described by the fast solution. Due to
the effect of a limb-darkened disk the mass loss rate increased in
an amount of $\sim 10\%$ while the terminal velocity is reduced 
about the same factor. Therefore, the limb darkening effect
should be considered always in the calculation of the hydrodynamics fast solution.

On the other hand, the influence of $f_{\mathrm{\,LD}}$ on the $\Omega$-slow and
$\delta$-slow solutions is minimal. The maximum difference
obtained in the velocity profile computed with uniformly bright and
limb-darkened disk radiation sources is less than 3 $km\,s^{-1}$
for the $\Omega$-slow solution, 7 $km\,s^{-1}$ for the $\delta$-slow
solution and 1.5 $km\,s^{-1}$ for the case when both parameter
$\delta$ and $\Omega$ are high (the $\Omega \delta$-slow solution).
Therefore, the limb darkening effect is negligible when computing
the wind parameters. However, rotational effects like the star's
oblateness  should be considered, since it modifies the wind in the
polar direction \citep[see][]{ara11} being much faster than the spherical
one. Moreover, the slow solutions predict even slower and denser flows
than the spherical ones.  

The influence of $f_{\mathrm{\,LD}}$ on radiation driven winds can be interpreted 
in terms of the resulting velocity profile. The mayor differences between the uniformly 
bright and limb-darkened finite disk correction factors are in
the region just above the stellar photosphere, as Figure \ref{fig1} shows.
In this region the velocity from all the models described in section \ref{results}
are small, however the value of the velocity gradient 
from the fast-solution is 5 to 10 times larger than the values from any slow-solution.  
Figure \ref{fig6} shows the normalized velocity gradient $dw/du$ as function  
of $u$ for the different types of solutions. Is this dependence on the velocity gradient,
specifically in the finite disk correction factor, that makes a significant difference in the 
terminal velocity and the mass loss rate {\it{only}} for the fast solution and not for the slow ones.

Concerning the WM-L relationship, the limb-darkened correction
factor has no effects. The increase produced by the fast solution 
on $\dot{M}$ is compensated by a similar decrease of $v_{\infty}$.  
Considering the importance of having a
theoretical  WM-L relationship for  B- and A- type supergiants the effect of the
star's oblateness and gravity-darkening should be explored, together
with the calculation of the synthetic line spectrum in order to
derive accurate wind parameters.

\begin{acknowledgements}

MC acknowledges financial support from 
Centro de Astrof\'{\i}sica de Valpara\'{\i}so and from CONICYT, Departamento de 
Relaciones Internacionales “Programa de Cooperaci\'on Cient\'ifica Internacional” 
CONICYT/MINCYT 2011-656. LC acknowledges financial 
support from the Agencia de Promoci\'on Cient\'{\i}fica y Tecnol\'ogica 
(BID 1728 OC/AR PICT 0885), from CONICET (PIP 0300), and the Programa 
de Incentivos G11/109 of the Universidad Nacional de La Plata, Argentina.
DR acknowledges financial support from CONICET (PIP 112200901000637).
\end{acknowledgements}

\appendix

\section{Partial derivatives of $f_{\mathrm{\,LD}}$}
\label{A}
In order to find the location of the singular point, we need to evaluate 
the singularity condition given by eq. \ref{2.6} and, then, impose the 
regularity condition given by eq. \ref{2.7}. To perfom this calculation 
we need to know all the partial derivatives of
$f_{\mathrm{\,LD}}\,(u,w,w')$; i.e., 
$\partial f_{\mathrm{\,LD}}/\partial u$, $\partial f_{\mathrm{\,LD}}/\partial w$ 
and $\partial f_{\mathrm{\,LD}}/\partial w'$.\\
Defining the following auxiliary variables:
\begin{eqnarray}
Z &=& w/w' \\
\lambda &=&u \,(u+Z)
\end{eqnarray}
Thus, in terms of $\lambda$, the finite disk correction factor for a uniformly 
bright spherical star $f_{\mathrm{D}}$ reads,
\begin{equation} 
f_{\mathrm{D}}\,(\lambda)=\frac{1}{(1+\alpha)}\,\frac{1}{\lambda}\, \left[1-\left(1-\lambda
\right)^{\,(1+\alpha)} \right], 
\end{equation} 
while the limb-darkened finite disk correction factor $f_{\mathrm{LD}}$ is:
\begin{equation}
f_{\mathrm{LD}}\,(\lambda)=\frac{\left(1-\lambda\right)^{\alpha }\,
   \left[\left(1-\lambda\right)^{-\alpha }+ \lambda\, (\alpha +1) 
   \,\, _2F_1\left(\frac{3}{2},-\alpha
   ,\frac{5}{2},\frac{\lambda }{\lambda
   -1}\right)+\lambda -1\right]}{2\,\lambda\,(\alpha
   +1)}
\end{equation}
Defining now $e\,(\lambda)=\partial f_{\mathrm{\,LD}}\,(\lambda)/ \partial \lambda $, we obtain:
\begin{eqnarray}
e\,(\lambda)& =& \frac{\left(1-\lambda\right)^{\alpha }}{4\,(\alpha +1)\,(\lambda -1)\, \lambda ^2} \times\, \left[ (2-(3 \alpha +5)\,\lambda )
   \left(1-\lambda\right)^{-\alpha} +\right. \nonumber \\
& &\left. +2\, (\lambda -1)\,(\alpha  \lambda+1)+(\alpha +1)\,\lambda\,(2 \alpha \lambda +3) \, {}_2F_1\left(\frac{3}{2},-\alpha
   ,\frac{5}{2},\frac{\lambda}{\lambda-1}\right) \right]
\end{eqnarray}
Therefore, all the partial derivatives can be calculated using the chain rule, getting,
\begin{equation} 
e\,(\lambda )=\frac{1}{2\,u + w/w'} \,\frac{\partial f_{\mathrm{\,LD}}}{\partial
u} = \frac{w'}{u} \,\frac{\partial f_{\mathrm{\,LD}}}{\partial w}=-\frac{w^{\prime\,2}}{u\, w} \, \frac{\partial f_{\mathrm{\,LD}}}{\partial w^{\prime}}.  
\end{equation} 

\bibliographystyle{apj}
\bibliography{citas}
\end{document}